\documentclass[10pt,a4paper]{article}
\usepackage[latin1]{inputenc}
\usepackage{amsmath}
\usepackage{amsfonts}
\usepackage{amssymb}
\usepackage{graphicx}
\usepackage{authblk}
\begin{document}

\title{\textbf{A novel radioguided surgery technique exploiting $\beta^{-}$ decays}}

\author[a]{E.~Solfaroli Camillocci}
\author[b]{G.~Baroni}
\author[c,d]{F.~Bellini}
\author[d]{V.~Bocci}
\author[c,d]{F.~Collamati} 
\author[e]{M.~Cremonesi}
\author[j]{E.~De~Lucia}
\author[f]{P.~Ferroli}
\author[d,g]{S.~Fiore}
\author[e]{C.~M.~Grana}
\author[h,d]{M.~Marafini}
\author[i,j]{I.~Mattei}
\author[d]{S.~Morganti}
\author[m]{G.~Paganelli}
\author[k,d]{V.~Patera}
\author[k,j]{L.~Piersanti}
\author[d]{L.~Recchia}
\author[a,c,d]{A.~Russomando}
\author[f]{M.~Schiariti}
\author[k,j]{A.~Sarti}
\author[k,d]{A.~Sciubba}
\author[d]{C.~Voena}
\author[c,d]{R.~Faccini}
\affil[a]{Center for Life Nano Science@Sapienza, 
Istituto Italiano di Tecnologia, Roma, Italy.}
\affil[b]{Dip. Elettronica, Informazione e Bioingegneria, Politecnico di Milano, Italy;}
\affil[c]{Dip. Fisica, Sapienza Univ. di Roma, Roma, Italy;}
\affil[d]{INFN Sezione di Roma, Roma, Italy;}
\affil[e]{Istituto Europeo di Oncologia, Milano, Italy;}
\affil[f]{Fondazione Istituto Neurologico Carlo Besta, Milano, 
Italy;}
\affil[g]{ENEA UTTMAT-IRR, Casaccia R.C., Roma, Italy;}
\affil[h]{Museo Storico della Fisica e Centro Studi e Ricerche 
`E.~Fermi', Roma, Italy;}
\affil[i]{Dipartimento di Fisica, Universit\`a Roma Tre, Roma, 
Italy;}
\affil[j]{Laboratori Nazionali di Frascati dell'INFN, Frascati, 
Italy;}
\affil[k]{Dip. Scienze di Base e Applicate per l'Ingegneria, 
Sapienza Univ. di Roma, Roma, Italy.}
\affil[m]{Department of Nuclear Medicine and Radiometabolic Unit,
Istituto Scientifico Romagnolo per lo Studio e la Cura dei Tumori, IRST-IRCCS, Meldola, Italy}
\maketitle

\begin{abstract}
  The background induced 
by the high penetration power of the $\gamma$ radiation 
is the main limiting factor of the current Radio-guided surgery (RGS). To partially mitigate it, a RGS with $\beta^+$-emitting radio-tracers has been suggested in literature.
 Here we propose the use of $\beta^-$-emitting radio-tracers  and $\beta^-$ probes and discuss the advantage of this method with respect to the previously explored ones:
the electron low penetration power 
allows for simple and versatile probes
and could extend RGS 
to tumours for which background originating from nearby healthy tissue 
makes $\gamma$ probes less effective.
We developed a $\beta^-$ probe prototype and  studied its performances on phantoms. By means of a detailed simulation we have also extrapolated the results to estimate the performances in a realistic case of meningioma, pathology which is going to be our first in-vivo test case.
A good sensitivity to residuals down to 0.1~ml  
can be reached within 1~s with an administered activity 
smaller than those for PET-scans thus
making the radiation exposure to medical personnel negligible. 
\end{abstract}

\begin{figure}[htbp] 
\centering
\includegraphics[width=\textwidth]{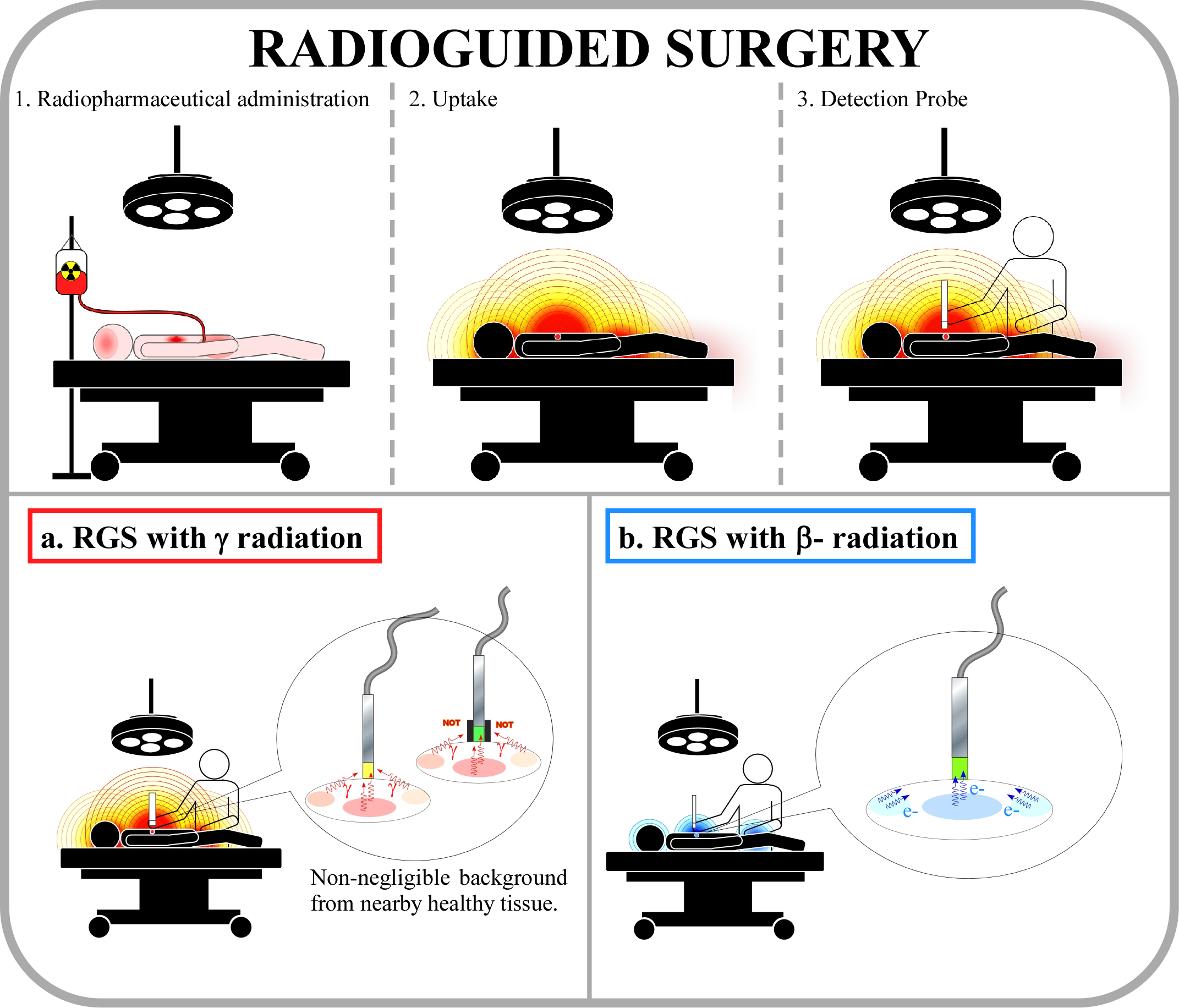}
\caption{{\bf The radioguided surgery technique (RGS)}. Steps of the procedure:
(1) a radio-labelled tracer is administered to the patient, before the surgery; 
(2) the emitting tracer is preferentially taken up by the tumour; 
(3) after the cancerous bulk removal,
the surgeon explores the lesion with a radiation detecting probe 
and looks for targeted tumour residuals in real time.
\newline
The bottom boxes show the effect of the proposed replacement of the $\gamma$-emitting tracers (a)
with electron-emitting tracers (b).
Due to the high penetration power of the photons,
in the first case a non-negligible background can be 
produced by the healthy organs close to the lesion, 
sometimes preventing the applicability of the technique. 
To mitigate this effect a shielding or active veto is applied 
(see inset of box a) thus making the probes cumbersome.
Electrons, instead, provide a clearer delineation
of radioactive tissue's margins 
allowing for a simple and compact probe
and requiring a smaller radio-pharmaceutical activity.}
\label{fig:RGS}
\end{figure}

The radio-guided surgery (RGS) is a surgical technique, 
first developed some 60 years ago, 
that enables the surgeon 
to evaluate the completeness of the tumoural lesion resection, 
while minimizing the amount of healthy tissue removed~\cite{RadioGuided}.
The impact of the RGS on the surgical management of cancer patients 
includes providing the surgeon with vital and real-time information
regarding the location and the extent of the lesion, 
as well as assessing surgical resection margins.
The technique (see Fig.~\ref{fig:RGS})
makes use of a radio-labelled tracer, 
preferentially taken up by the tumour  
to mark the cancerous tissue from the healthy organs, 
and a probe (for a review see~\cite{IntrProbes}), 
sensitive to the emission released by the tracer, 
to identify in real time the targeted tumour loci. 
The radio-pharmaceutical is administered to the patient 
just before the surgery. 

Current clinical applications of the RGS are: 
radio-immuno-guided surgery (RIGS) for colon cancer, complete 
sentinel-node mapping for malignant melanoma and breast cancer, 
and detection of parathyroid adenoma and bone tumours 
(such as osteoid osteoma).

 Established methods make use of a combination of
a $\gamma$-emitting tracer with a
$\gamma$ radiation detection probe (see~\cite{GammaProbes} and references therein).
Since $\gamma$ radiation can traverse large amounts of tissue,
any uptake of the tracer in nearby healthy tissue
represents a non-negligible background, 
often preventing the usage of this technique. 

To mitigate this effect it was suggested in literature the use of  $\beta^+$ decaying tracers~\cite{beta+technique}. The emitted positrons in fact have a limited penetration and their detection is local. Nonetheless, positrons annihilate with electrons in the body and produce $\gamma$s with an energy of 511 keV: the background persists and actually increases in energy. The improvement with respect to the use of pure $\gamma$ emitters is that a dual system can be devised where the background can be measured separately and subtracted from the observed signal. This approach has been studied in preclinical tests~\cite{beta+preclinical} but it is not yet in use in the clinical practice. The largest limitations range from the time needed to identify a residual, the dimensions of the probes and the dose absorbed by the medical personnel. 

A better solution to the current limits of  RGS would be to eliminate the background from $\gamma$ radiation. This study suggests therefore the use of  pure $\beta^{-}$  radiation that
penetrates only a few millimetres of tissue with essentially no $\gamma$ contamination, being the \textit{bremsstrahlung} contribution, with a 0.1\% emission probability, negligible. 
Furthermore, a $\beta^{-}$ probe, 
detecting electrons and operating with lower background, 
provides a clearer delineation of margins of the radioactive tissue, 
requires administration of a radio-pharmaceutical with a lower activity
and it is smaller and easier to handle in the surgical environment.
The lower absorbed dose, 
together with the short range of electrons,
implies an almost negligible radiation exposure to medical personnel 
and therefore a larger number of RGSs per year for the surgeon.

The idea of detecting $\beta^{-}$ emission was first 
proposed at the very beginning of the development of 
RGS~\cite{ProbeHistory}. 
However, it was soon abandoned due to the particular 
phenomenology of this emission. 
The short electron free-path demanded for low invasiveness detectors 
to be used intra-operatively in contact with the tissues and 
sensitive enough to absorbed dose normally accepted in nuclear medicine. 
Both requirements were not satisfied by instruments 
and materials available at the time, 
while are easily reachable with modern technologies.

 In particular, we report on the developments in the field of the radio-tracers and in the development and test of a specific intra-operative probe device that make the  RGS with $\beta^{-}$ decays feasible.
\subsection*{Results}
\paragraph{The clinical case.}
There are several possible applications of this technique, many of which probably have not yet been thought of. First, complete removal of tumour is critical for brain tumours, where a relapse is particularly dangerous and where 
other RGS techniques are limited by the high uptake of the brain. 
Next, 
particular attention should be paid to the complete resection 
of the main tumour 
and of infected lymph-nodes in the case of pediatric tumours 
where life expectancy is long. 
Such tumours are typically abdominal and 
therefore probe signals are blinded by background from kidneys, liver and bladder.  
Finally, there are abdominal tumours in adults, 
like non-palpable metastases in liver and insulinoma, 
that would profit from RGS.

Testing this novel technique with most of the above-mentioned 
clinical cases requires first to identify and test an 
appropriate $\beta^{-}$-emitting tracer. 
We then focused on a brain tumor for which such a tracer already exists.  
Based on our experience with the Peptide Receptor Radionuclide Therapy 
(PRRT)~\cite{IEO1, IEO2, IEO3}, the meningioma has been chosen:
this brain tumor is particularly receptive to 
synthetic somatostatin analogues, such as DOTATOC, 
that can be labelled with the $\beta^{-}$ emitting $^{90}$Y. 
Furthermore, the same tracer can be marked with $^{68}$Ga, thus allowing for an estimate of 
uptake prior to surgery with a Positron Emission Tomography (PET) exam. Finally, it was recently shown that with an appropriate SPECT device also the brehmsstrahlung emission of  the $^{90}$Y
can be use to verify the receptivity~\cite{Y90-SPECT}.

\paragraph{The $\beta^{-}$ detecting probe.}
As far as the probe design is concerned, 
the choice of the materials and the readout electronics is driven 
by the need to maximize the sensitivity to electrons
while minimizing the sensitivity to photons.
Para-terphenyl, or p-terphenyl, was adopted as electron detector 
after a detailed study~\cite{PTerf} 
due to its high light yield and low density,
with consequent low sensitivity to photons.

A first prototype of the $\beta^{-}$ probe was developed 
(see Fig~\ref{fig:probe})
and tested with phantoms
using $^{90}$Y diluted in a physiological saline solution
(see Methods for the experimental set-up
and Supplementary Information for more details on the test results).
We explored the range 
of activity concentration from 22 to 5~kBq/ml
to be compared with the 20~kBq/ml value 
estimated by analyzing PET 
DICOM images obtained 
administering 3~MBq/kg of $^{68}$Ga-DOTATOC 
to patients affected by meningioma.
We therefore studied the probe performance
close and below the activity range achieved
for diagnostic investigation. 

We explored the first prototype response with four 
phantoms simulating cancerous residuals with different 
topologies (see Supplementary Fig. S1).
The 0.1~ml volume phantom, referred to as "RESIDUAL",
has dimensions compatible with residuals well identified 
with the nuclear magnetic resonance. 
To check the effect of the phantom depth on the probe response
and resolution in distinguishing the residual edge, the other three cylindrical phantoms have the same activity concentration, footprint 
(13~mm$^2$) but different heights: 1, 2, and 3~mm referred to as "H1", "H2, "H3" respectively.

A 1.5~mm step blind automated scan over the phantom engraved surface was performed when the activity concentration reached 16~kBq/ml.  
Even by limiting the acquisition time to 1~s per step the test demonstrated that all the residuals would be identifiable in absence of background from nearby organs.
The signal from the tumor residuals was studied at maximum and minimum activity concentrations, 22 and 5~kBq/ml, and the observed rates are reported in Tab.~\ref{tab:phantom}. The electronic noise was found to be negligible.
Background rates, 
dominated by the signal coming from the uptake of nearby healthy tissue, 
were estimated using a full simulation of the system with FLUKA program~\cite{FLUKA}, 
based on the above mentioned DICOM images showing that 
the nearby healthy tissue uptake is approximately 10 times smaller 
than the tumour.

From the signal and background rates, 
taking into account the Poisson fluctuations of the measured counts, 
we computed the false-positive (FP) and the false-negative (FN) rates 
for a given time interval measurement with the probe.
We estimated (see Tab.~\ref{tab:phantom}) 
the time needed to achieve FP$\approx 1\%$ and FN$<5\%$
and found it to be 1~s for 22~kBq/ml activity concentration 
and up to 10~s in the case of a concentration as low as 5~kBq/ml. 
Further improvements of the probe are needed to reduce such time at a level compatible with surgical environment and administer down to 5~kBq/ml.

The effect of an off-axis measurement has been verified 
by moving the probe away from the phantom centre.  
The test showed that the rate is reduced by a factor 2 when moving the probe 0.5~mm far from the phantom edge (see Supplementary Fig. S4). This allows us to conclude that the probe distinguishes the residual edge.
This test also demonstrated the lateral shielding effectiveness of this prototype making the probe insensitive to electrons coming from the sides 
hence increasing the tumor-spotting capability.  This test also shows that the probe is insensitive of long-range radiation that could be caused by the activity of the $^{90}$Y in nearby samples.

\paragraph{Radiation exposure to patient and medical personnel}
One of the key points of this technique is the low irradiation of the patient 
and above all of the medical staff.
As previously mentioned, we estimated that 
a 22~kBq/ml activity concentration, 
corresponding to administering approximately 3~MBq/kg,
would be more than sufficient for the technique to be effective.

From Ref.~\cite{IEO3} we obtained that the total body dose absorbed by a 70~kg patient after administration of 210~MBq of $^{90}$Y-DOTATOC is approximately 21~mSv, while the effective dose is about 70~mSv.
This corresponds to approximately two Computed Tomography (CT) exams and 
therefore there is room for improvement in reducing the required activity.

To evaluate the dose absorbed by the surgeons 
we simulated a set-up similar to the common situation 
of an operating room. 
Both activities of neoplastic cells and normal tissue 
were taken into account, 
according to the ratios obtained 
from the aforementioned studies on PET images.
The equivalent dose absorbed by the surgeon's hands
computed for RGS with a $^{90}$Y emitting tracer
is expected to be smaller than 1~$\mu$Sv/hour.
The corresponding value for established RGS with $^{99m}$Tc radio-label tracer is 24~$\mu$Sv/hour as estimated with the simulation, which was consistent with Ref.~\cite{RadioGuided}. Similarly, the total-body dose to the surgeon is 0.13~$\mu$Sv/hour with the $\beta^-$ RGS here proposed, 
to be compared with approximately 6~$\mu$Sv/hour of the $^{99m}$Tc RGS.

\subsection*{Discussion}

The radio-guided surgery 
represents a very useful surgical adjunct 
in those cases where a complete resection, 
intended as full enhancing mass removal, 
is crucial both for recurrence-free survival and 
the overall survival of the patients, 
particularly for those tumours 
where the surgical mass removal is the only possible therapy.

We are proposing a radical change in the paradigm of this technique:
the use of $\beta^{-}$ radiation instead of $\gamma$ 
or $\beta^+$ radiation. 
The reduced penetration power of electrons should allow 
the extension of this technique to clinical cases 
that would be otherwise prevented 
by the presence of nearby healthy organs taking up the tracer. 

This novel approach allows to develop a simple and compact probe which, 
detecting electrons and operating with low radiation background, 
provides a clearer delineation of margins of radioactive tissue
and requires a smaller radio-pharmaceutical activity 
to detect tumour remnants compared to traditional RGS approaches.
Moreover, due to the lower absorbed dose and the short range of electrons, the radiation exposure to the patient and medical personnel becomes almost negligible.
 Such considerations apply also in the comparison with the $\beta^+$ RGS, since in the latter the $\gamma$ background is in any case present, albeit subtracted with the dual-probe approach. Furthermore, measurements performed in absence of background, as in the case of the proposed $\beta^-$ RGS, require by definition  shorter read-out times than measurements with background subtraction. Finally, the need to duplicate the probe for the background estimation makes $\beta^+$ probes intrinsically less compact. 

We have presented pre-clinical tests of a prototype probe supporting the above statements. From these measurements we extrapolated with a detailed simulation the expected performance with meningioma, representing our test clinical case. Nonetheless, the actual uptake on the margins of the lesion,  the impact of tissue between the probe and the residual and the effect of the nearby blood is something to be estimated in clinical tests.
Once the feasibility of such technique will be demonstrated with meningioma, it will be possible to extend it to other clinically relevant cases, eventually together with the development of specific radio-tracers.

\subsection*{Methods}
\label{sec:Methods}

\begin{figure}[htbp]
\centering
 \includegraphics[width=7cm]{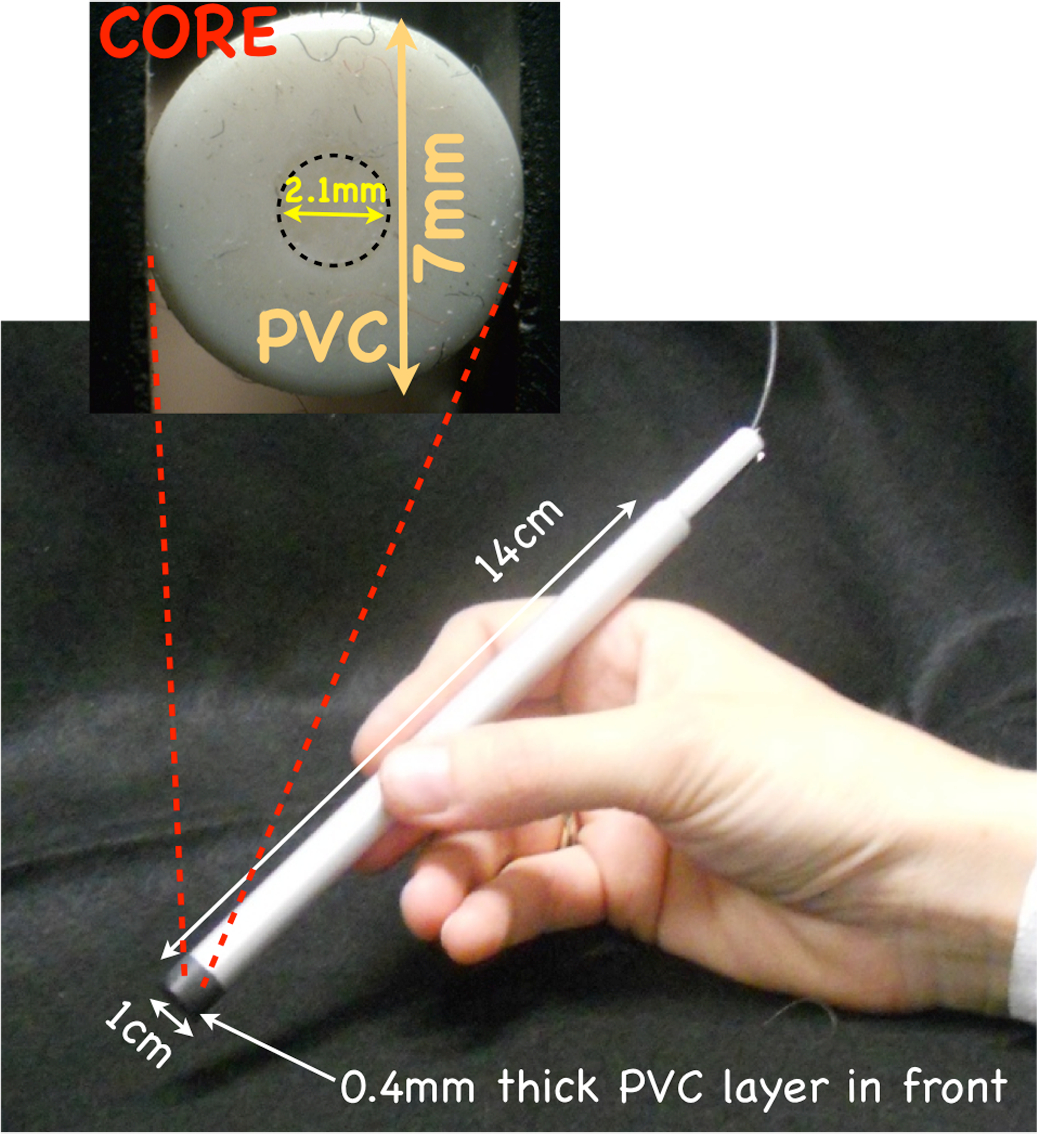}
 \caption{\textbf{First prototype of the intraoperative
 $\beta^-$ probe.} 
 The core is a cylindrical scintillator 
 (diameter 2.1~mm, height 1.7~mm)
 of poli-crystalline p-terphenyl.
 A ring of PVC wraps the scintillator and 
 shields it against radiation coming from the sides. 
 The device is encapsulated inside an easy-to-handle aluminum body 
 as protection against mechanical stress
 and it is protected against light by a thin PVC layer.
}
\label{fig:probe}
\end{figure}

\paragraph{The probe prototype.}
The core of the probe (see Fig.~\ref{fig:probe})  is a cylindrical scintillator 
(diameter 2.1~mm, height 1.7~mm)
of poli-crystalline para-terphenyl doped 
by 0.1\% in mass of diphenylbutadiene.
Para-terphenyl was adopted, after 
a detailed study~\cite{PTerf}, 
due to its high light yield 
(3.5 times larger than typical organic scintillators),
non-hygroscopic property and low density.
The scintillator is shielding against radiation coming from the sides by wrapping it with a 7~mm external diameter ring of PVC.
The device is encapsulated inside an easy-to-handle aluminum body,
as protection against mechanical stress,
and has a blinding 0.4~mm-thick black PVC front end cap.
The scintillation light is transported to a photo-multiplier tube 
(PMT, Hamamatsu H10721-210), through an optical fibre,
and read out by a portable custom electronics 
with wireless connection to a PC or tablet. 
This prototype is compatible with a standard sterile covering 
of sub-millimetric film for surgical environment.  Also the very low bias required by the PMT, 5V, makes the device easily portable.

\paragraph{Experimental set-up of the $^{90}$Y test.} The phantoms 
are cylindrical vessels, 
engraved at a fixed radius on a PMMA (Plexiglass$^{\texttt{TM}}$) disk 
mounted on a rotating table,
as shown in supplementary Fig. S1.
Their dimensions are listed in Tab.~\ref{tab:phantom}
and are known with a precision of 10~$\mu$m.

During the test, 
the probe was fixed in vertical position over the disk 
and the phantoms were moved under the probe
by rotating the disk with a step motor, controlled via PC.
The relative position of the sensitive head of the probe and the 
 phantom 
was known with a precision of 10~$\mu$m.
Two probe-disk distances were studied: 
50 and 150~$\mu$m. 
At the optimal working point
(PMT gain $\sim$10$^6$ and
signal readout threshold of 21~mV),
the background rate due to 
the photo-multiplier tube dark current and casual events
was measured to be 0.2 counts per second (cps).
It was checked to be stable when operating at room-temperature
and with ambient light conditions.

\paragraph{Estimate of false positive and false negative rates.}
The rate of false positive and false negative (FP and FN) is computed from the signal ($\nu_s$) and background ($\nu_b$, mostly due to tracer uptake in healthy tissue) rates estimated, 
for each phantom, as described in the Results section. 
For a given value of the probe acquisition time ($t_{daq}$), the number of  signal counts from the tumour and the background is distributed according to a poisson distribution with mean $\mu_s=\nu_s t_{daq}$ and $\mu_b=\nu_b t_{daq}$, respectively.
Given the minimum number of signal counts needed to flag a positive identification $th$, FP is computed as the fraction of times the background would yield a positive signal:
\begin{equation}
FP=1-\sum_{N=0}^{th-1}{\cal{P}}_{\mu_b}(N)
\end{equation}
where ${\cal{P}}_{\mu}(N)$ indicates the Poisson probability to have $N$ if the mean is $\mu$.
Similarly FN is the fraction of times a tumour residual would not yield a signal
\begin{equation}
FN=\sum_{N=0}^{th-1}{\cal{P}}_{\mu_s}(N) .
\end{equation}

To determine the minimum acquisition time reported in Tab.~\ref{tab:phantom}, 
FN and FP are computed in a grid of $t_{daq}$ and $th$ (see Supplementary Fig.S5)
and the smallest value of $t_{daq}$ for which FN$<5\%$ and FP$\approx 1\%$ is determined.

\section*{Acknowledgements}
We would like to thank the Institute for Nuclear Medicine of the Policlinico A. Gemelli, Universita' Cattolica del Sacro Cuore of Rome for the stimulating discussions.

\section*{Author contributions}
R.F., A.Sc., S.F. ideated the project and supervise it by identifying the materials and their supply; 
S.M., E.S.C., A.R., F.C., F.B., M.M., A.S. worked on the design and construction of the probe prototype, automated test station implementation, data analysis and interpretation, probe optimization; 
R.F., E.S.C., S.M., F.C. prepared the manuscript;
F.C., V.P., L.P., I.M., C.V., E.D.L. implemented the Monte Carlo simulation aimed to compare different materials and to understand the probe behavior;  
G.B., P.F., M.S. provided the medical background and defined the clinical case; 
G.P., C.M.G., M.C. provided the nuclear medicine knowledge and supplied DICOM data to study use cases;
V.B., L.R.  developed and optimized custom electronics for the probe readout; 
all authors discussed the results and implications and commented on the manuscript at all stages.

\paragraph{Competing financial interests}
F.B., F.C., E.D.L., S.F., M.M., I.M., V.P., L.P., A.Sa., A.Sc., C.V. and R.F
are listed as inventors on an Italian patent application 
(RM2013A000053) 
entitled
``Utilizzo di radiazione $\beta^-$ per la identificazione intraoperatoria 
di residui tumorali e la corrispondente sonda di rivelazione" 
dealing with the implementation of an
intra-operative $\beta^-$ probe for radio-guided surgery
according to the results presented in this paper.

\begin{table}[htbp]
\begin{center}
\begin{tabular}{|l|c|c|c|c|c|c|c|}
\hline 
Phantom & Diameter & Height & Volume & Rate (cps) & T (s) & Rate (cps) & T (s)\\
		& (mm) & (mm) & (ml) & 22 kBq/ml & 22 kBq/ml & 5 kBq/ml & 5 kBq/ml \\ 
\hline 
Residual & 6 & 3.5 & 0.10 & 31.6 & 1 & 6.6 & 2 \\
H1       & 4 & 1   & 0.01 & 12.4 & 2 & 2.6 & $>$10 \\
H2       & 4 & 2   & 0.02 & 17.7 & 1 & 3.7 & 4 \\
H3       & 4 & 3   & 0.04 & 20.1 & 1 & 4.2 & 4 \\
\hline
\end{tabular}
\end{center}

\caption{\textbf{Results of the test on the first probe prototype.} 
The probe was tested on phantoms sized as possible tumour residuals of interest, 
filled with $^{90}$Y in saline solution to simulate the situation 
after bulk meningioma removal.
The rates measured with two different $^{90}$Y activity concentrations (22 and 5~kBq/ml) are reported (Rate).   
The minimal acquisition time (T) needed to detect tumour residuals
with a false-negative probability $<$5\% 
and a false-positive probability $\sim$1\% was estimated extrapolating the laboratory test results to a real case by means 
of a detailed simulation.
}
 \label{tab:phantom}
\end{table}

\section*{Supplementary Information}

\subsection*{Probe test in laboratory}

\begin{figure}[h]
\centering
  \includegraphics[width=0.7\textwidth]{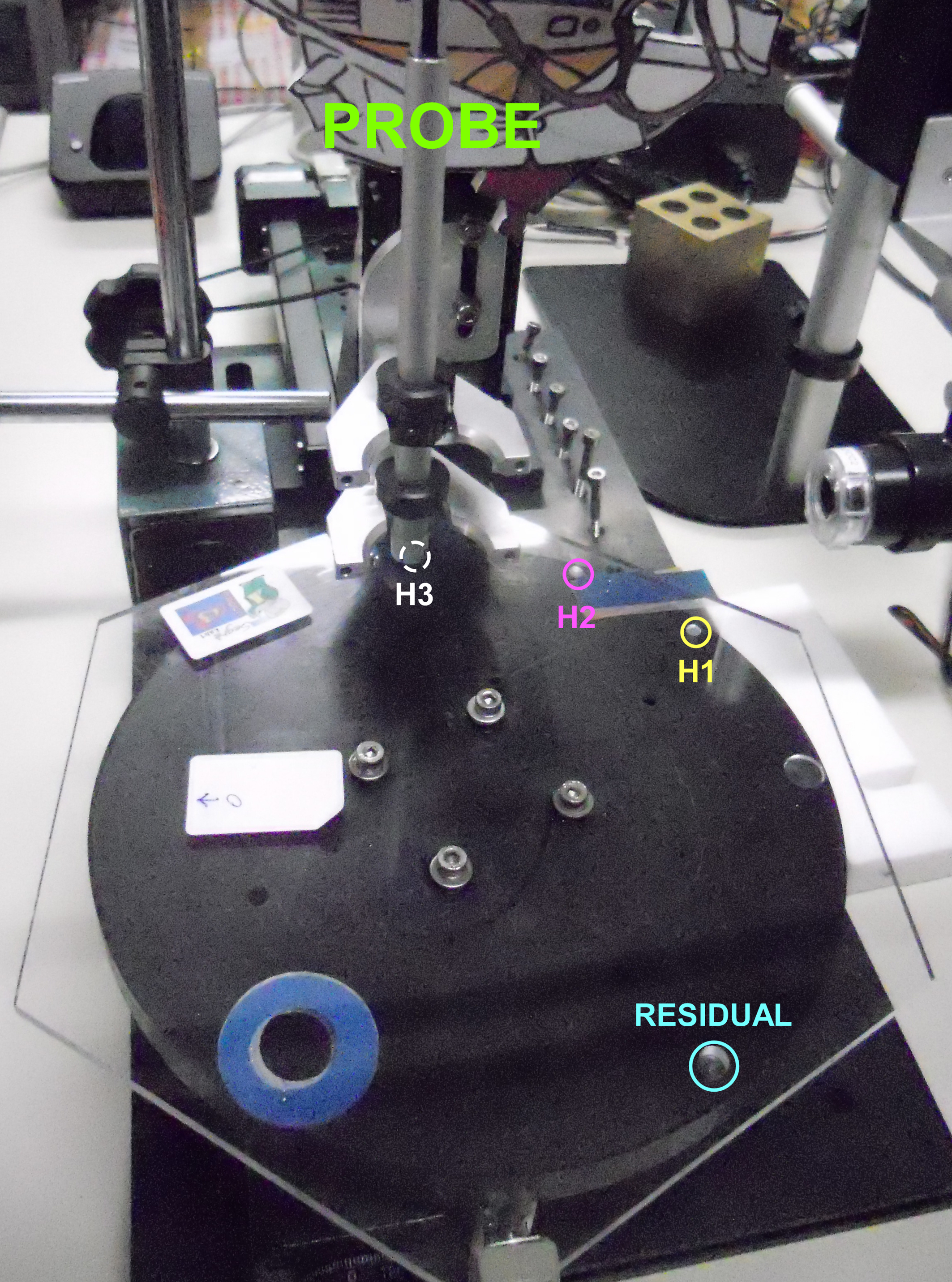}
 \caption[*]{\textbf{Experimental set-up.} 
 The probe and the automated disk with phantoms for the $^{90}$Y tests.
}
 \label{fig:TestDISK}
\end{figure}

\paragraph{Supplementary Discussion on the probe test results.}
We observed that the probe prototype is able 
to detect samples of millimetric dimensions and 
different topologies 
in the whole activity concentration range suitable for diagnostic investigation 
(5-22~kBq/ml).
The counts-per-second (cps) measured by the probe on any phantom 
decrease, as expected, 
 with the $^{90}$Y activity 
according to the radionuclide decay law,
as shown in Supplementary Fig.~\ref{fig:RESIDrate} 
for the RESIDUAL phantom.
\begin{figure}[h]
\centering
 \includegraphics[width=\textwidth]{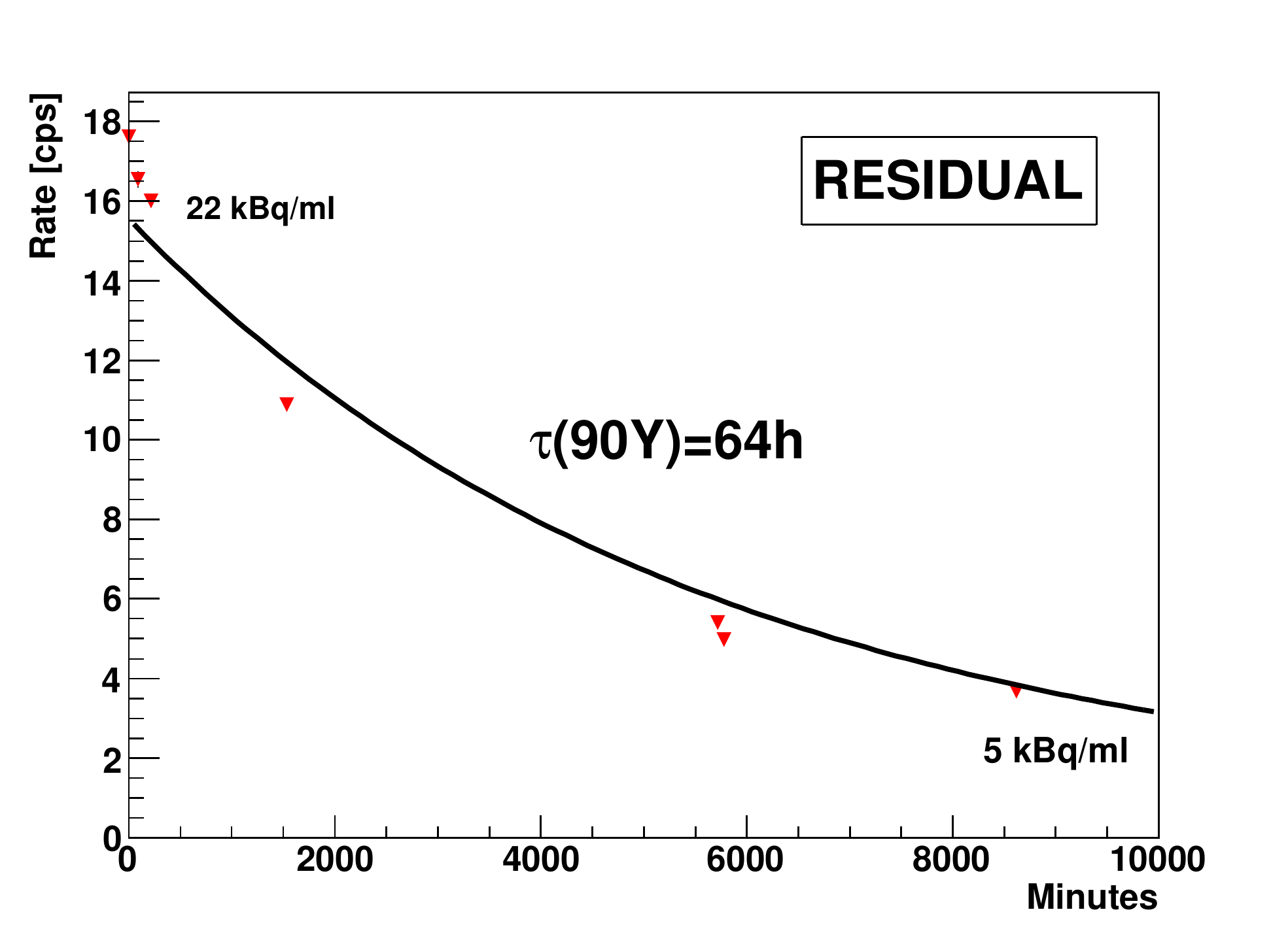}
 \caption[*]{\textbf{Counts-per-second observed during 1 week 
 by the probe over the RESIDUAL phantom filled with $^{90}$Y 
 in physiological saline solution.}
 As expected, the rate decreases as the $^{90}$Y activity.  The superimposed line is the result of a fit with lifetime fixed to the one of $^{90}$Y. 
 }
 \label{fig:RESIDrate}
\end{figure}

To test the detection efficiency of the probe
and its sensitivity to different phantom topologies,
we performed a blind scan
simulating the surgeon exploring the area to look for residuals.
Fixing the probe position over the phantom disk
with a random offset with respect to the phantom positions,
we rotate the motorized table by a complete turn with 1 degree step angle
(corresponding to 1.5~mm step along the circumference)
recording a 1~s-long rate measurement at each step. 
As shown in Supplementary Fig.~\ref{fig:PROBEturn} 
all four phantoms were clearly detected
at the 16~kBq/ml $^{90}$Y activity concentration.  Furthermore, the absence of a signal when the probe was located in between the samples demonstrates the insensitivity to any long-range radiation that could be produced by the activity of the phantoms.

\begin{figure}[h]
\centering
 \includegraphics[width=\textwidth]{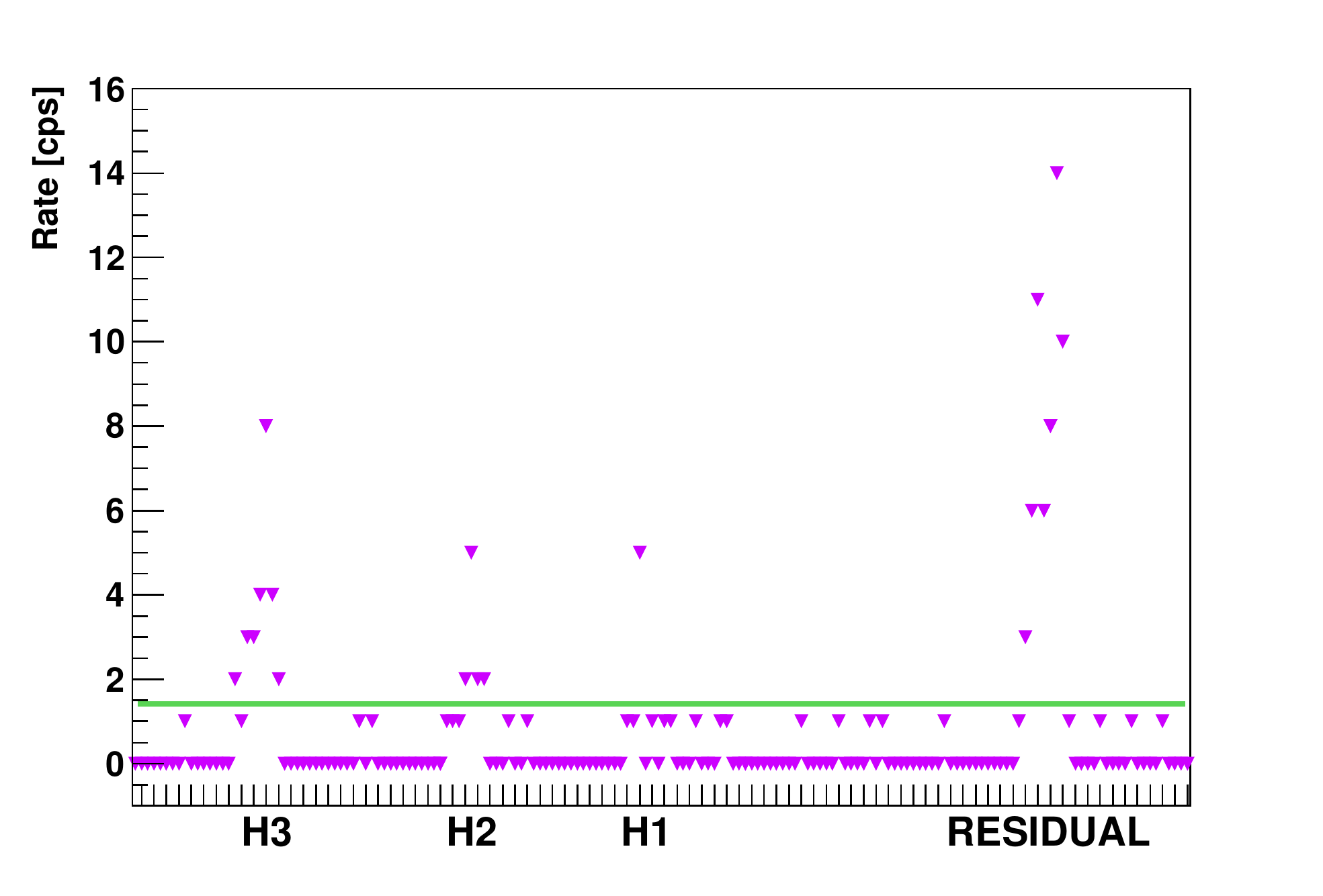}
 \caption{\textbf{Blind scan simulating surgeon exploring the area 
 to look for hot spots.} 
 The counts were measured during a complete turn of the phantom disk
 with step of 1 degree and acquisition time of 1~s for each step. 
 The rate increases when the probe is over a radioactive volume. Considering two counts per second sufficient to spot a residual, all samples would have been detected.
 The $^{90}$Y activity concentration was 16~kBq/ml and
 the distance between the probe and the phantoms was 50~$\mu$m.
 }
 \label{fig:PROBEturn}
\end{figure}

The effect on the probe sensitivity of different phantom's footprints and 
of the radioactive volume thickness was also tested.  
As shown in Supplementary Fig.~\ref{fig:rateEdge},
moving the probe away from the phantoms
results in a decrease in rate, which is halved when the centre of the probe is 0.5~mm from the edge of the sample.
This is true for any phantom shape demonstrating that 
the depth of the sample does not affect the sensitivity of the device to the phantom's edge, making it possible 
to identify the real dimension of the active area.
\begin{figure}[h]
\centering
 \includegraphics[width=\textwidth]{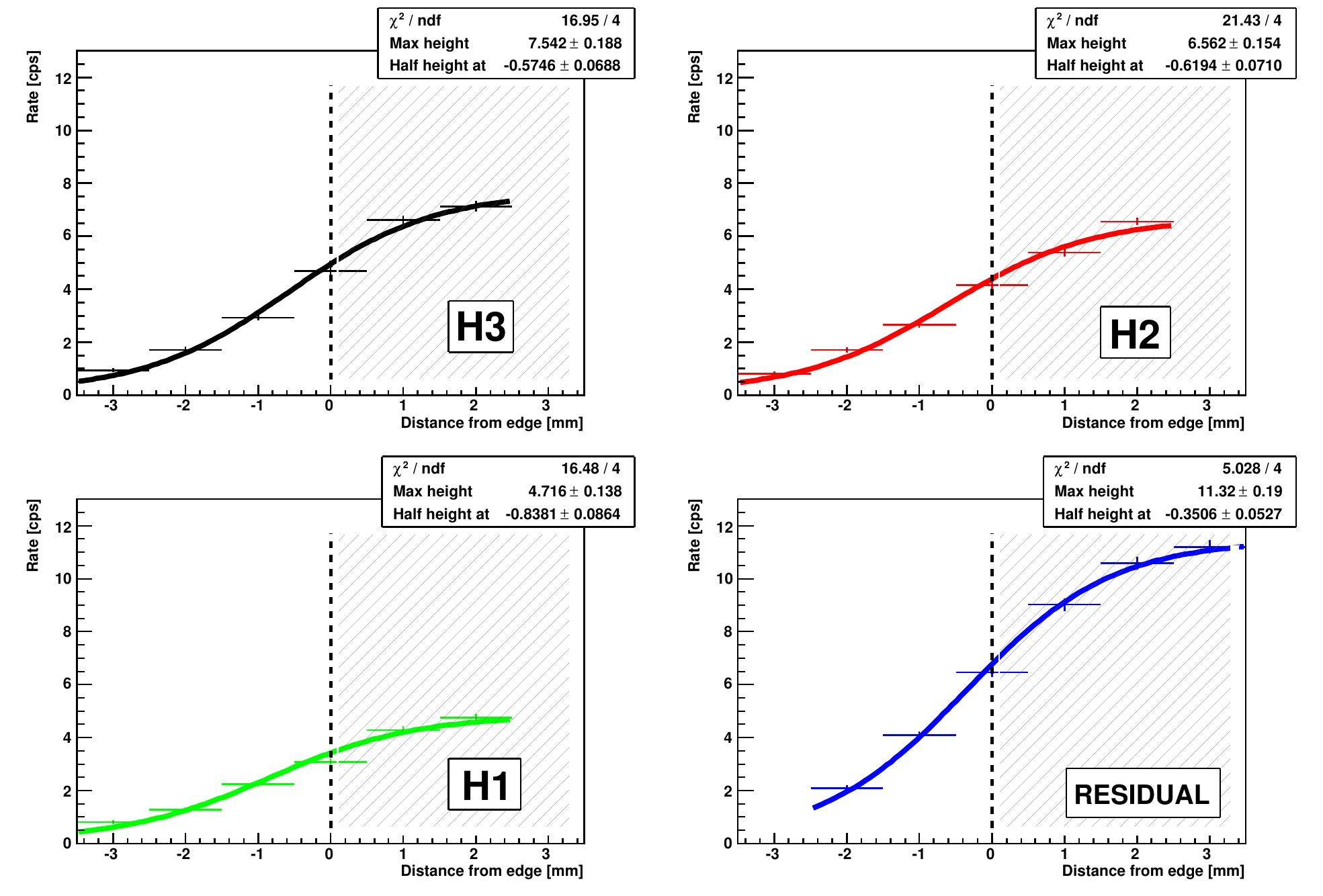}
 \caption{\textbf{Counts-per-second measured moving the probe 
 away from the centre of the phantoms with step of 1~mm.} 
 In all cases the cps decreases 
 reaching half of the central value when
 the probe is about 0.5~mm far from the radioactive volume edge.
 The $^{90}$Y activity concentration was 16~kBq/ml and
  the distance between the probe and the phantoms was 50~$\mu$m.
}
 \label{fig:rateEdge}
\end{figure}

 In conclusion  we verified that by administering a $^{90}$Y activity similar to those used for diagnostic purposes, the probe prototype is able to identify millimetric radioactive volumes.
 It can detect the real dimension of the phantoms
providing a fast response.

\subsection*{Determination of the minimal acquisition time.}
\begin{figure}[h]
\centering
 \includegraphics[width=\textwidth]{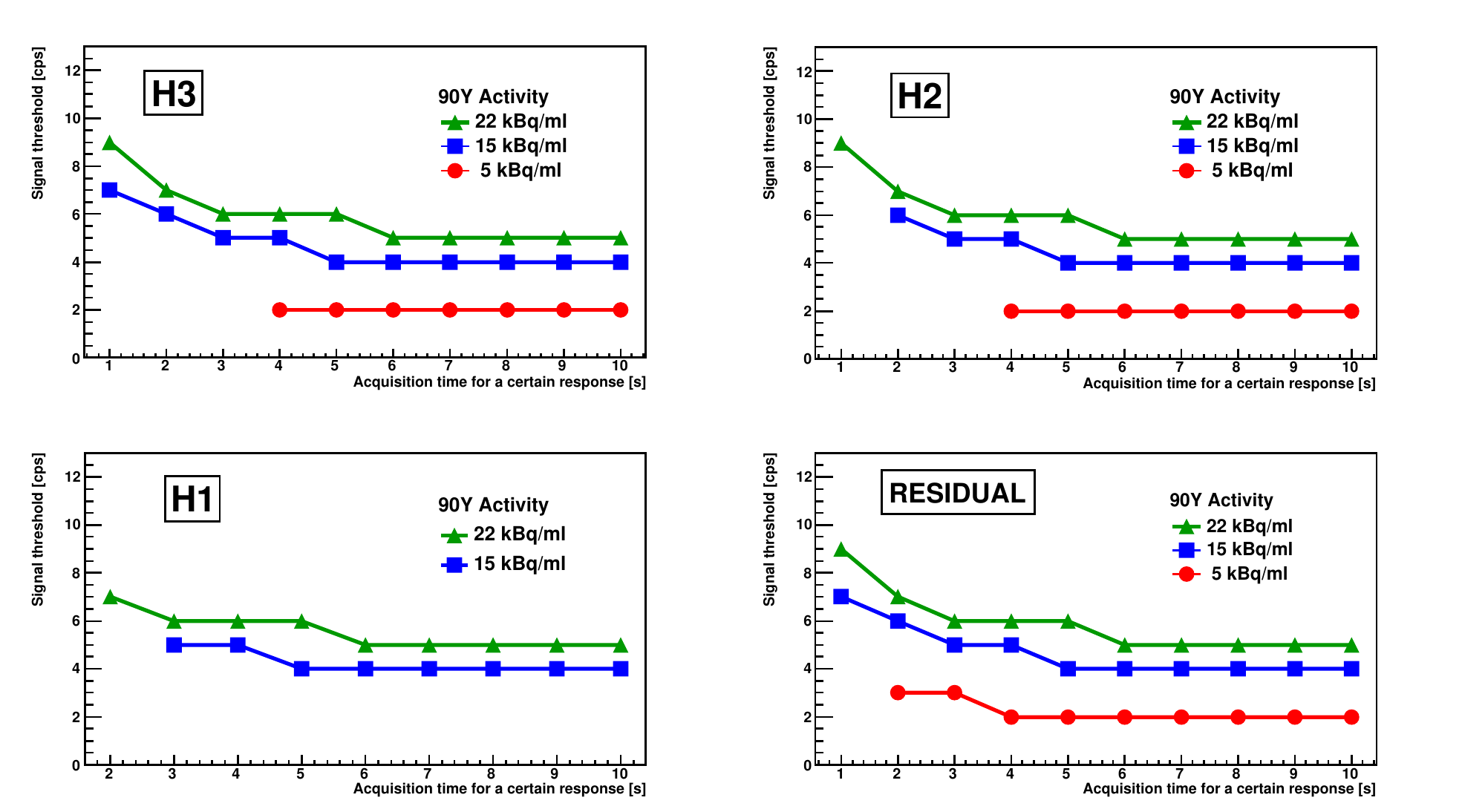}
 \caption{\textbf{Determination of the minimal acquisition time.}
 As described in the Methods section,
  FN and FP depend on the acquisition time ($t_{daq}$)
 and the minimum signal required to claim a positive evidence ($th$).
 By requiring a false-positive probability of $\sim$1\%
 and a false-negative probability $<$5\%
 the allowed working points for $t_{daq}$ and $th$ are described.
 The results for the four phantoms under study are compared
 considering administration of radio-pharmaceutical
 with three different activity values.
 These curves were estimated in the first test medical case, 
 the meningioma marked with $^{90}$Y-DOTATOC.
 } 
 \label{fig:PROBEwpoint}
\end{figure}

Supplementary Fig.~\ref{fig:PROBEwpoint} shows the combination 
of $th$ and $t_{daq}$ with FP$\approx$1\% and FN$<5$\%. 
These curves were used, as described in the Methods section, 
to estimate the minimum acquisition time needed at each activity.

\begin{thebibliography}{9}

\bibitem{RadioGuided}
{\it Radioguided Surgery: A Comprehensive Team Approach.} [Mariani, G. , Giuliano, A. E, Strauss, H.W.  (eds) ] (Springer, New York, 2006).

\bibitem{IntrProbes}
Hoffman,E.J. Tornai,M.P. Janecek,M. Patt,B.E. and Iwanczyk, J.S.
Intraoperative probes and imaging probes.
\textit{Eur. J. Nucl. Med.} {\bf 26,} 913-935 (1999)

\bibitem{GammaProbes}

Tsuchimochi M. and Hayamaand K. Intraoperative gamma cameras for radioguided surgery: Technical characteristics, performance 
parameters, and clinical application, {\textit{ Phys. Med.}}  {\bf 29,} 126-38 (2013) 

\bibitem{beta+technique}
 Hickernell T. S. et al. Dual detector Probe for surgical Tumor Staging, { \textit {J. Nucl. Med.}}, {\bf 29,} 1101 (1988);
 Daghighian, F. et al. Intraoperative beta probe: A device for detecting tissue labeled with positron or electron emitting isotopes during surgery  {\textit{Med. Phys.}}  {\bf21,} 153 (1994); Raylman, R. R. and  Wahl R. L. A fiber-optically coupled positron-sensitive surgical probe, {\textit{J. Nucl. Med.}}, {\bf 35,} 909 (1994); Bonzom S. An Intraoperative Beta Probe Dedicated to Glioma Surgery: Design and Feasibility Study, {\textit{  IEEE Trans. Nucl. Sci. }} {\bf54,} 1 (2007). 
 
\bibitem{beta+preclinical}
Raylman R. R. et al. Fluorine-18-fluorodeoxyglucose-guided breast cancer surgery with a positron-sensitive probe: Validation in preclinical studies {\textit {J. Nucl. Med.}}, {\bf 36,} 1869 (1995);
Zervos E. E. et al. 18F-Labeled Fluorodeoxyglucose Positron Emission Tomography-Guided Surgery for Recurrent Colorectal Cancer: A Feasibility Study {\textit{ J. of Surg. Res.}} {\bf97,} 9 (2001);
Bogalhas,F. et al.
Development of a positron probe for localization and
excision of brain tumours during surgery, {\textit{ Phys. Med. Biol.}} {\bf 54,} 4439 (2009);
Gonzales S.J. et al An analysis of the utility of handheld PET probes for the intraoperative localization of malignant tissue, {\textit{ J. Gastrointest. Surg.}} {\bf15,} 358-55 (2011); Singh,B. et al. A hand-held beta imaging probe for FDG.
\textit{Ann. of Nucl. Med.} \textbf{27,} 203-208 (2013).

\bibitem{ProbeHistory}
Selverstone,B. Solomon,A.K. Sweet,W.H. 
Location of brain tumors by means of radioactive phosphorous.
\textit{The J. of the American Med. Ass.} \textbf{140,} 277-278 (1949)

\bibitem{IEO1} 
Cremonesi,M. Ferrari,M. Bodei,L. Tosi,G. and Paganelli,G.
Dosimetry in peptide radionuclide receptor therapy: a review.
\textit{J. Nucl. Med.} \textbf{47,} 1467-1475 (2006)

\bibitem{IEO2}
Bartolomei,M. et al.
Peptide receptor radionuclide therapy with $^{90}$Y-DOTATOC 
in recurrent meningioma.
\textit{Eur. J. Nucl. Med. Mol. Imaging} \textbf{36,} 1407-1416 
(2009)

\bibitem{IEO3}
Cremonesi,M. et al. Dosimetry for treatment with radiolabelled somatostatin 
analogues. A review.
\textit{Q. J. Nucl. Med. Mol. Imaging} \textbf{54,} 37-51 (2010)

\bibitem{Y90-SPECT}
Fabbri, C. et al.   Quantitative evaluation on $^{90}Y-$DOTATOC PET and SPECT imaging by phantom acquisitions and clinical applications in locoregional and systemic treatments. {\textit Q. J. Nucl. Med. Mol. Imaging} \textbf{56,} 522-8 (2012)

\bibitem{PTerf}
Angelone,M. et al. 
Properties of para-terphenyl as detector for alpha, beta and 
gamma radiation.
 {\textit {arXiv:1305.0442}} (2013)

\bibitem{FLUKA}
G. Battiston, et. al
The FLUKA code: Description and benchmarking.
\textit{{AIP Conf. Proc.}} \textbf{896,} 31-49 (2006)

\end{thebibliography}
\end{document}